# Data Processing Benchmarks

Jérôme Darmont

Université de Lyon (Laboratoire ERIC), France

## INTRODUCTION

Performance measurement tools are very important, both for designers and users of database systems, whether they are aimed at On-Line Transaction Processing (OLTP) or On-Line Analysis Processing (OLAP). Performance evaluation is useful to designers to determine elements of architecture, and more generally to validate or refute hypotheses regarding the actual behavior of a system. Thus, performance evaluation is an essential component in the development process of well-designed and scalable systems, which is nowadays of primary importance in the context of cloud computing. Moreover, users may also employ performance evaluation, either to compare the efficiency of different technologies before selecting a software solution or to tune a system.

Performance evaluation by experimentation on a real system is generally referred to as benchmarking. It consists in performing a series of tests on a given system to estimate its performance in a given setting. Typically, a database benchmark is constituted of two main elements: a data model (conceptual schema and extension) and a workload model (set of read and write operations) to apply on this dataset, with respect to a predefined protocol. Most benchmarks also include a set of simple or composite performance metrics such as response time, throughput, number of input/output, disk or memory usage, etc.

The aim of this article is to present an overview of the major families of state-of-the-art data processing benchmarks, namely transaction processing benchmarks and decision support benchmarks. We also address the newer trends in cloud benchmarking. Finally, we discuss the issues, tradeoffs and future trends for data processing benchmarks.

**BACKGROUND**

**Transaction processing benchmarks**

In the world of relational database benchmarking, the Transaction Processing Performance Council (TPC) plays a preponderant role. The mission of this non-profit organization is to issue standard benchmarks, to verify their correct application by users, and to regularly publish the results of performance tests. Classical TPC benchmarks all share variants of a classical business database (customer-order-product-supplier) and are only parameterized by a scale factor that determines the database size (e.g., from 1 GB to 100 TB).

The TPC benchmark for transactional databases, TPC-C (TPC, 2010a), has been in use since 1992. It is specifically dedicated to OLTP applications, and features a complex database (nine types of tables bearing various structures and sizes), and a workload of diversely complex transactions that are executed concurrently. The performance metric in TPC-C is throughput, in terms of transactions. TPC-C was complemented in 2007 by TPC-E (TPC, 2010b), which simulates a brokerage firm with the aim of being representative of more modern OLTP systems that those modeled in TPC-C. TPC-E's principles and features are otherwise very similar to TPC-C's.

There are currently very few alternatives to TPC-C and TPC-E, although some benchmarks have been proposed to suit niches in which there is no standard benchmark. For instance, OO7 (Carey et al, 1993) and OCB (Darmont & Schneider, 2000) are object-oriented database benchmarks modeling engineering applications (e.g., computer-aided design, software engineering). However, their complexity makes both these benchmarks hard to understand and implement. Moreover, with objects in databases being more commonly managed in object-relational systems nowadays, object-relational benchmarks such as BUCKY (Carey et al., 1997) and BORD (Lee et al., 2000) now seem more relevant. Such benchmarks focus on queries implying object identifiers, inheritance, joins, class and object references, multivalued attributes, query unnesting, object methods, and abstract data types. However, typical object navigation is considered already addressed by object-oriented benchmarks and is not taken into account. Moreover, object-relational database benchmarks have not evolved since the early 2000's, whereas object-relational database systems have.

Finally, XML benchmarks aim at comparing the various XML storage and querying solutions developed since the late nineties. From the early so-called XML application benchmarks that implement a mixed XML database that is either data-oriented (structured data) or document-oriented (in general, random texts built from a dictionary), XBench (Yao et al., 2004) stands out. XBench is indeed the only benchmark proposing a true mixed dataset (i.e., data *and* document-oriented) and helping evaluate all the functionalities offered by XQuery. FlexBench (Vranec & Mlýnková, 2009) also tests a large set of data characteristics, but also proposes query templates that allow modeling multiple types of applications. Finally, Schmidt et al. (2009) and Zhang et al. (2011) propose benchmarks that are specifically tailored for testing logical XML model-based systems, namely native XML and XML-relational database management systems, respectively.

**Decision support benchmarks**

Since decision-support benchmarks are currently a de facto subclass of relational benchmarks, the TPC again plays a central role in their standardization. TPC-H (TPC, 2013) has long been the only standard decision-support benchmark. It exploits a classical product-order-supplier database schema, as well as a workload that is constituted of twenty-two

SQL-92, parameterized, decision-support queries and two refreshing functions that insert tuples into and delete tuples from the database. Query parameters are randomly instantiated following a uniform law. Three primary metrics describe performance in terms of power, throughput, and a combination of power and throughput.

However, TPC-H's database schema is not a star-like schema that is typical in data warehouses. Furthermore, its workload does not include any OLAP query. TPC-DS (TPC, 2012) now fills in this gap. Its schema represents the decision-support functions of a retailer under the form of a constellation schema with several fact tables and shared dimensions. TPC-DS' workload is constituted of four classes of queries: reporting queries, ad-hoc decision-support queries, interactive OLAP queries, and extraction queries. SQL-99 query templates help randomly generate a set of about five hundred queries, following non-uniform distributions. One primary throughput metric is proposed in TPC-DS, which takes both query execution and the data warehouse maintenance into account.

There are, again, few decision-support benchmarks out of the TPC, but with TPC-DS having been under development for almost eight years, alternative data warehouse benchmarks were proposed. Published by the OLAP council, a now inactive organization founded by OLAP vendors, APB-1 (OLAP Council, 1998) was the first of them and actually predates TPC-DS. APB-1 has been intensively used in the late nineties. However, APB-1 is very simple and rapidly proved limited to evaluate the specificities of various activities and functions.

Thus, more elaborate alternatives were proposed, such as DWEB (Darmont et al., 2007), which can be parameterized to generate various ad-hoc synthetic data warehouses and workloads that include typical OLAP queries, and SSB (O'Neil et al, 2009), which is based on TPC-H's database remodeled as a star schema and features a query workload that provides both functional and selectivity coverage.

Eventually, benchmarks also fill in niches that are not covered by the TPC. As SSB, XWeB (Mahboubi & Darmont, 2010) derives from TPC-H, but proposes of a test data warehouse based on a unified reference model for XML warehouses and its associate XQuery decision-support workload. RTDW-bench (Jedrzejczak et al., 2012) is also based on TPC-H. It is designed for testing the ability of a real-time data warehouse to handle a transaction stream without delay, given an arrival rate. Bär and Golab (2012) also propose a benchmark for stream data warehouses that measures the freshness of materialized views. Finally, a couple of benchmarks are even more specific (and unrelated from TPC-H), e.g., Spadawan (Lopes Siqueira et al., 2010), which allows performance evaluation of specific, complex operations in spatial data warehouses, and BenchDW (Triplet & Butler, 2013), which targets biological data warehouses and particularly focuses on performance metrics, with twenty-two different metrics such as documentation quality, accuracy and response time.

**Cloud benchmarks**

In the timely context of cloud computing and big data processing and analysis, benchmarking needs are as high as ever to compare parallel processing capability or infrastructure scalability and adaptability. Although no standard benchmark has emerged yet, some proposals have been made both by industry and academia.

MalStone (Open Cloud Consortium, 2009) is a benchmark for assessing data intensive parallel processing. It features MalGen, a synthetic data generator that produces large datasets generated probabilistically following specified distributions. In the same line, SWIM (Chen et al., 2013) is an open source benchmark that enables rigorous performance measurement of MapReduce systems. SWIM contains suites of workloads of thousands of jobs, with complex data, arrival, and computation patterns, and therefore provides workload-specific optimizations. SWIM is currently integrated with Hadoop. By contrast, YCSB (Cooper et al., 2010) is a framework that focuses on data, and more specifically on performance evaluation of key-value stores. YCSB defines several metrics and workloads to measure system behavior in different situations, or the same system when using different configurations.

OLTP-Bench (Curino et al., 2012) is the first true benchmarking framework designed for cloud transactional database systems as a service. OLTP-Bench actually features a set of existing micro-benchmarks (i.e., designed to test one very specific aspect of performance, e.g., ResourceStresser), popular benchmarks (e.g., TPC-C) and real-world applications (e.g., Wikipedia).

Finally, PRIMEBALL (Ferrarons et al., 2013) aims at providing a real-life context to cloud data warehouse benchmarking. Its authors provide the specifications of a fictitious news site hosted in the cloud that is to be managed by the framework under analysis, together with several objective use case scenarios and measures for evaluating system performance.

**ISSUES AND TRADEOFFS IN DATA PROCESSING BENCHMARKS**

Gray (1993) defines four primary criteria to specify a "good" benchmark:
1. relevance: the benchmark must deal with aspects of performance that appeal to the largest number of potential users;
2. portability: the benchmark must be reusable to test the performances of different DBMSs;
3. simplicity: the benchmark must be feasible and must not require too many resources;
4. scalability: the benchmark must adapt to small or large computer architectures.

In their majority, existing benchmarks aim at comparing the performances of different systems in given experimental conditions. This helps vendors in positioning their products relatively to their competitors', and users in achieving strategic and costly software choices based on objective information. These benchmarks invariably present fixed database sche-mas and workloads. Gray's scalability factor is achieved through a reduced number of parameters that mainly allow varying the database size in predetermined proportions. It is notably the case of the unique scale factor parameter that is used in all TPC benchmarks.

This solution is simple (still according to Gray's criteria), but the relevance of such benchmarks is inevitably reduced to the test cases that are explicitly modeled. For instance, the typical customer-order-product-supplier that is adopted by the TPC is unsuitable to many application domains other than management. This leads benchmark users to design more or less elaborate variants of standard tools, when they feel these are not generic enough to fulfill particular needs. Such users are generally not confronted to software choices, but to architectural choices or performance optimization tradeoffs within a given system or family of systems. In this context, it is essential to multiply experiments and test cases, and a monolithic benchmark is of reduced relevance.

To enhance the relevance of benchmarks aimed at system designers, two solutions are possible. The first one is to design an ad-hoc benchmark for a particular application, e.g., RTW-bench, Spadawan and BenchDW, for real-time, spatial and biological data warehouses, respectively. But then, the benchmark's application span is necessarily quite narrow. The alternative is to resort to benchmark generators, also called tunable or generic benchmarks, such as OCB, DWEB or FlexBench, which help generate various database or workload configurations, and thus allow experiments to be performed in various conditions. However, this approach is mechanically detrimental to simplicity, which is a primordial criterion. It is thus necessary to devise benchmark generators that to not sacrifice simplicity too much.

**FUTURE TRENDS**

The previous section showed that classical transaction and decision-oriented benchmarks are well-established. However, benchmarking in the cloud faces a new paradigm and must measure new features. Thus, in addition to Gray's (1993) criteria for building a good benchmark, Folkerts et al. (2012) propose that the quality criteria that are commonly accepted by the benchmarking community must be revisited when benchmarking in the cloud.

Although the cloud inherits from a long legacy of distributed systems, important issues are unique to the cloud. For instance, the concept of elasticity applied to data management may translate in the ability to dynamically bring in new data sources to meet emerging needs (Pedersen, 2010). Thus, cloud benchmark databases must be dynamic. With respect to security, specific issues appeared in the new framework of the cloud, e.g., cloud provider or subcontractor espionage, cost-effective defense of availability, uncontrolled mashups... (Chow et al., 2009). Such features are important to assess, for security is the top concern of cloud users and would-be users. Finally, the economic model of the cloud is fundamentally new. Instead of a costly initial investment, pay-as-you-go models allow users to pay a small amount per use, e.g., of a dataset, in return for a one-time advantage (Pedersen, 2010). Thus, cost must also be a key criterion when benchmarking cloud solutions.

All these aspects of cloud computing are in need of specific benchmarking, a new trend that is currently emerging, e.g., with some of PRIMEBALL's metrics, which not only target transaction performance, but also storage costs and data consistency, for instance. Bermbach et al. (2013) further advocate for a standard comprehensive benchmark for quantifying the consistency guarantees of eventually consistent storage systems.

Eventually, Folkerts et al. (2012) insist that executing a benchmark in a complex environment such as the cloud necessitates at least as much effort as designing it in the first place. Currently existing cloud data processing benchmarks are presumably one step beyond in this perspective. Thus, the involvement of a major actor on the benchmarking scene, such as the TPC, would certainly help standardize cloud benchmarking processes and tools.

# CONCLUSION

Benchmarking is a small field, but it is nonetheless essential to database research and industry. It serves both engineering and research purposes, when designing systems or validating solutions; and marketing purposes, when monitoring competition and comparing commercial products.

Benchmarks might be subdivided in three classes. First, standard, general-purpose benchmarks such as the TPC's do an excellent job in evaluating the global performance of systems. They are well-suited to software selection by users and marketing battles by vendors, who try to demonstrate the superiority of their product at one moment in time. However, their relevance drops for some particular applications that exploit database models or workloads that are radically different from the ones they implement. Ad-hoc benchmarks are a solution. They are either adaptations of general-purpose benchmarks, or specifically designed benchmarks. Designing myriads of narrow-band benchmarks is not time-efficient, though, and trust in yet another new benchmark might prove limited in the database community. Hence, the last alternative is to use generic benchmarks that feature a common base for generating various experimental possibilities. The drawback of this approach is that parameter complexity must be mastered, for generic benchmarks to be easily apprehended by users. In conclusion, before starting a benchmarking experiment, users' needs must be carefully assessed so that the right benchmark or benchmark class is selected, and test results are meaningful.

It is nonetheless clear that the TPC plays a primordial role in the data benchmarking community, not only by issuing standards, but also by structuring and leading the community, e.g., by organizing the annual Technology Conference on Performance Evaluation and Benchmarking (TPCTC). This event does not only promote the TPC's activity, but also greatly encourages industrial and academic advances in the field of performance evaluation and benchmarking, whether they are related to the TPC or TPC benchmarks or not.

**ADDITIONAL READINGS**

**TERMS AND DEFINITIONS**

Benchmark: A standard program that runs on different systems to provide an accurate measure of their performance.

Synthetic benchmark: A benchmark in which the workload model is artificially generated, as opposed to a real-life workload.

Database benchmark: A benchmark specifically aimed at evaluating the performance of Database Management Systems (DBMSs) or DBMS components.

Database model: In a database benchmark, a database schema and a protocol for instantiating this schema, *i.e.*, generating synthetic data or reusing real-life data.

Workload model: In a database benchmark, a set of predefined read and write operations or operation templates to apply on the benchmark's database, following a predefined protocol.

Performance metrics: Simple or composite metrics aimed at expressing the performance of a system.

Cloud benchmarking: Use of cloud services in the respective (distributed) systems under test (Folkerts et al., 2012).